\documentclass[useAMS,usenatbib,a4paper,referee]{mn2e}
\usepackage{graphicx,times}
\usepackage{amsmath}
\usepackage{longtable}

\title[Linear Growth of Structure]
  {The Linear Growth of Structure in the $R_{\rm  h}=ct$ Universe}
  \author[Fulvio Melia]
    {Fulvio Melia\thanks{John Woodruff Simpson Fellow. Email: fmelia@email.arizona.edu} \\
     Department of Physics, The Applied Math Program, and Department of Astronomy, The University of Arizona, AZ 85721, USA}

\begin{document}


%

\maketitle

\begin{abstract}
We use recently published redshift space distortion
measurements of the cosmological growth rate, $f\sigma_8(z)$, to
examine whether the linear evolution of perturbations in the
$R_{\rm h}=ct$ cosmology is consistent with the observed development
of large scale structure. We find that these observations favour
$R_{\rm h}=ct$ over the version of $\Lambda$CDM optimized with
the joint analysis of {\it Planck} and linear growth rate data, particularly 
in the redshift range $0< z < 1$, where a significant curvature in the
functional form of $f\sigma_8(z)$ predicted by the standard model---but
not by $R_{\rm h}$$=$$ct$---is absent in the data. When $\Lambda$CDM is
optimized using solely the growth rate measurements, however, the two 
models fit the observations equally well though, in this case, the
low-redshift measurements find a lower value for the fluctuation
amplitude than is expected in {\it Planck} $\Lambda$CDM. Our results 
strongly affirm the need for more precise measurements of $f\sigma_8(z)$
at all redshifts, but especially at $z < 1$.
\end{abstract}

\section{Introduction}
The large-scale structure revealed by the distribution of galaxies
is believed to have formed through a process of gravitational instability,
starting with primordial fluctuations in the early Universe. But while
self-gravity amplifies perturbations in the cosmic fluid, cosmic expansion
suppresses them. Their growth rate therefore depends rather sensitively on
the dynamical expansion of the Universe and may be used to discriminate
between different models. Measurements of the growth rate tend to focus on
infall motions associated with condensing regions, with peculiar velocities
largely correlated with the local gravitational potential. Galaxies trace
these motions, carrying an imprint of the changing growth rate as the
Universe evolves (Peacock et al. 2001; Viel et al. 2004; Jain and Zhang 2007; Ross et al. 2007;
da Angela et al. 2008; Guzzo et al. 2008; Song and Koyama 2009; Song and Percival 2009;
Davis et al. 2011; Hudson and Turnbull 2012; Macaulay et al. 2013; Alam et al. 2016).

A principal statistical technique used to measure the growth rate is based on
the redshift-space distortion (RSD) created by the galaxies' peculiar velocities
(Kaiser 1987). Specifically, maps produced with distances inferred from
redshifts in spectroscopic galaxy surveys show that the galaxy distributions
are anisotropic due to the fact that the redshifts contain components from both
the smooth Hubble flow and the peculiar velocities of the infalling matter.
As long as one can reliably separate these two contributions to the redshift,
one may thereby extract a history of the build-up of structure.

The use of RSD, however, is  practical primarily before non-linear effects
begin to emerge, where measurements yield information on both the matter
over-density and the peculiar velocities of galaxies. In the linear regime,
the problem is typically reduced to solving a second-order differential equation
for the time-dependent fluctuations, from which one may then infer their growth rate.
Thus, although objects suitable for this work can in principle include individual
galaxies, clusters of galaxies, and superclusters, their density contrasts
($\delta\rho/\rho$) today are, respectively, $\sim 10^6$, $\sim 10^3$, and
$\sim O(1)$. These estimates assume a critical density $\rho_c\equiv 
3c^2H_0^2/8\pi G\sim 10^{-29}$ g cm$^{-3}$, and that about $10^{12}$
$M_\odot$ of galactic matter are contained within a $\sim 30$ kpc region.
Clusters typically contain fewer than $10^3$ galaxies, while superclusters
have tens of thousands of galaxies. In $\Lambda$CDM, $\rho\sim (1+z)^3$ 
during the matter-dominated era, so galaxies and clusters presumably
ceased growing linearly, i.e., grew with $\delta\rho< 1$, at redshifts
$\sim 100$ and $\sim 10$, respectively. These structures are non-linear in 
the local neighborhood. The linear-growth analysis described in this paper
therefore tends to address the formation of super-clusters, which could
have grown linearly over the redshift range $z < 2-3$. This is the
approach we shall follow in this paper, and therefore focus on surveys 
relevant to these large structures, including 2dFGRS (Peacock et al. 2001) 
and VVDS (Guzzo et al. 2008).

By now, the development of linear perturbation theory is quite mature---at least
within the context of the standard model ($\Lambda$CDM), in which dark energy
corresponds to a cosmological constant $\Lambda$. Comparative tests using the
growth of linear structure have been carried out between $\Lambda$CDM and an
assortment of other cosmologies, principally those based on extensions to general
relativity involving higher-order curvature terms or extra dimensions
(Wetterich 1995; Amendola 2000; Dvali et al. 2000; Carroll et al. 2004; Capozziello et al. 2005).
However, measurements of the linear growth rate have not yet been used to test the
$R_{\rm h}=ct$ universe (Melia 2007; Melia and Shevchuk 2012; Melia 2016a, 2017), another
FRW cosmology, which has thus far been shown to fit many other kinds of data
better than $\Lambda$CDM. (A brief summary of these previously published
results is provided in \S~3.1 below.) The principal aim of 
this paper is to address this deficiency.

We are especially motivated to carry out this analysis by recent comparative studies
using the Alcock-Paczy\'nski test (Alcock and Paczy\'nski 1979), based the changing
ratio of angular to spatial/redshift size of (presumed) spherically-symmetric source
distributions with distance (Melia and L\'opez-Corredoira 2016). The use of this 
diagnostic, with newly acquired measurements of the anisotropic distribution of 
BAO peaks from SDSS-III/BOSS-DR11 at average redshifts $\langle z\rangle=0.57$ 
and $\langle z\rangle=2.34$, has allowed us to determine the geometry of the 
Universe with unprecedented accuracy. Previous applications of the galaxy 
two-point correlation function to measure a redshift-dependent scale that could
be used to determine the ratio of angular (i.e., transverse) size to
redshift (i.e., radial) size were limited by the need to disentangle the acoustic
scale in redshift space from redshift distortions from internal gravitational
effects (L\'opez-Corredoira 2014). A major limitation of this process
was that inevitably one had to pre-assume a particular cosmological model, 
or adopt prior parameter values, in order to estimate the possible
confusion between the true cosmological redshift interval from one edge
of the cluster to the other and the contribution to this redshift width from
these internal gravitational effects. Unfortunately, the wide range of 
possible distortions for the same correlation-function shape resulted in 
very large errors associated with the BAO peak position and hence the
inferred acoustic scale (often in the $\sim 20-30\%$ range).

This situation has improved significantly over the past few years with 
(1) the use of reconstruction techniques (Eisenstein et al. 2007; 
Padmanabhan et al. 2012) that enhance the quality of the galaxy two-point 
correlation function, and (2) the use of Ly-$\alpha$ and quasar observations
to more precisely determine their auto- and cross-correlation functions, 
allowing the measurement of BAO peak positions to better than $\sim 4\%$ 
accuracy (Cuesta et al. 2016). 
The most recent determination of $y(z)$ has been based on the use
of three BAO peak positions: the measurement of the BAO peak position 
in the anisotropic distribution of SDSS-III/BOSS DR12 galaxies at 
$\langle z\rangle=0.32$ and $\langle z\rangle=0.57$ (Cuesta et al. 2016), 
in which a technique of reconstruction to improve the signal/noise ratio 
was applied; and the self-correlation of the BAO peak in the Ly-$\alpha$ 
forest in the SDSS-III/BOSS DR11 data at $\langle z\rangle=2.34$ 
(Padmanabhan et al. 2012), plus the cross-correlation of the BAO peak of 
QSOs and the Ly-$\alpha$ forest in the same survey (Font-Ribera et al. 2014).

With these new measurements, the use of the Alcock-Paczy\'nski diagnostic
(Melia and L\'opez-Corredoira 2016) has shown that the current concordance 
($\Lambda$CDM) model is disfavoured by the BAO data at $2.6\sigma$. They
instead show that the $R_{\rm h}=ct$ model has a probability 
$\sim 0.68$ (i.e., consistent with 1) of being correct. Measurements of the 
linear growth rate also critically depend on the Alcock-Paczy\'nski effect, 
so the observations considered in this paper provide an invaluable, 
complementary, set of data with which to test the $R_{\rm h}=ct$ cosmology.

In \S~2 of this paper, we derive the necessary formalism for studying the
time evolution of linear fluctuations in this model, which reduces to solving
a second-order differential equation, though with some important differences
compared with its counterpart in $\Lambda$CDM. A contextual background for
$R_{\rm h}=ct$ is provided in \S~3, where we also solve the growth equation
as a function of redshift, and describe the observables, specifically the 
volume-delimited variance $\sigma_8(z)$ of the fluctuations and its 
corresponding growth function. The standard model is analyzed in \S~4, and 
we end with a discussion and conclusion in \S\S~5 and 6.

\section{Relativistic Perturbation Theory}
The equations describing the growth of linear perturbations in $\Lambda$CDM
are well known so, for this model, we will simply adopt the key results from previous
work and summarize these in \S~4 below. In the case of $R_{\rm h}=ct$, however,
it is essential to begin with relativistic perturbation theory, which we now describe.
We will assume that small inhomogeneities emerge into the
semi-classical universe out of the Planck regime (Melia 2016b)
with an essentially scale-free distribution (see below). The subsequent
development of structure is usually assumed to have progressed through
a series of steps, each corresponding to a particular, dominant
component in the cosmic fluid, eventually leading to the formation
of stars and galaxies. To study the growth of perturbations in an
otherwise smooth background, it is helpful to know (i) the relative
abundance of non-relativistic baryonic (b) and cold-dark matter (cdm);
(ii) the contribution to the total energy density $\rho$ and pressure
$p$ from all the components in the cosmic fluid; and (iii) the spectrum
and type (i.e., adiabatic or isothermal) of primeval density
fluctuations.

Newtonian theory, as a limiting approximation to general relativity (GR),
is only applicable within the Hubble radius $R_{\rm  h}\equiv c/H(t)$,
where the effects of spacetime curvature are small. In the $R_{\rm h}=ct$
universe, the quantum fluctuations always have a wavelength (or scale)
$\lambda$ smaller than $R_{\rm h}$, and the perturbations that grow
from these early seeds never stretch in size beyond this gravitational
horizon. But the full GR theory is nonetheless still necessary when
we are dealing with perturbations in relativistic matter, such as
would occur in a baryonic fluid coupled to the radiation field, or
in a fluid that includes energetic neutrinos.

We consider perturbations about the background solution for the density
and pressure, labeled with subscript $0$, such that
\begin{equation}
\rho=\rho_0+\delta\rho\;,
\end{equation}
and
\begin{equation}
p=p_0+\delta p\;,
\end{equation}
where $\delta\rho$ and $\delta p$ are the perturbed first-order variables
that generally depend on the spatial coordinates $x^i$ as well as the
time ($x^0=ct$). (In this paper, we denote the spatial coordinates with
indices $i,j,k$, while Greek indices refer to all four coordinates,
with metric signature $[+,-,-,-]$.) In the linear regime, we have
$|\delta\rho|\ll\rho$, and similarly for all the other first-order
quantities. Different perturbative modes therefore evolve independently
of each other and may be treated separately.

For simplicity, we consider only barotropic fluids with $p=p(\rho)$ and,
following convention, we use a dimensionless density variable $\delta\equiv
\delta\rho/\rho_0$, etc. in all the equations. We define the background
equation-of-state parameter
\begin{equation}
w\equiv {p_0\over\rho_0}\;,
\end{equation}
which includes the contributions to $\rho_0$ and $p_0$ from all the
components in the cosmic fluid, and the adiabatic sound speed, $v_s$,
within the fluctuation, where $\partial_\alpha \delta p=v_s^2\partial_\alpha 
\delta\rho$, so that
\begin{equation}
v_s^2\equiv {dp\over d\rho}\;,
\end{equation}
which depends principally on the perturbed quantities $\delta p$ and $\delta\rho$.

In deriving the dynamical equations for the evolution of the perturbations,
one must carefully distinguish between the different roles played by the
background quantities $\rho_0$ and $p_0$, and the fluctuations $\delta\rho$
and $\delta p$. The former dominate the evolution of the Hubble constant $H(t)$
in (cosmic) time $t$, while the local growth is heavily influenced by the
gravitational potential associated with the fluctuations themselves. Of
course, Einstein's Field Equations contain a single stress-energy tensor
$T^{\mu\nu}$ encompassing all of the sources (i.e., $\rho=\rho_0+\delta\rho$,
$p=p_0+\delta p$) but, as we shall see, once the dynamical equations are
linearized, some terms depend predominantly on $\rho_0$ and $p_0$, while
others contain only the perturbed amplitudes. The equations we will use
allow for the possible evolution of a matter perturbation embedded
within an otherwise smooth background, such as we would encounter in
a radiation-dominated, or dark energy dominated, universe. In
every case, however, the expansion rate is always driven by
$\rho\approx\rho_0$ and $p\approx p_0$.

In the relativistic treatment of fluctuation growth, we perturb
both the spacetime metric and the stress-energy tensor $T^{\mu\nu}$
representing the sources. The details of how one linearizes the Einstein
Field Equations have appeared in many previous publications, and
we refer the reader to some of these excellent works
(Weinberg 1972; Landau and Lifshitz 1975; Peebles 1980; Press and
Vishniac 1980; Kolb and Turner 1990; Padmanabhan 1993; Peebles 1993; Coles and 
Lucchin 1995; Peacock 1999; Liddle and Lyth 2000; Tsagas et al. 2008). We follow the covariant
Lagrangian approach, employing locally-defined quantities, and we
derive their evolution along the worldlines of {\it comoving} observers.
There is therefore a slight difference between the proper time $\tau$
defined at each spacetime point, and the cosmic time $t$, since the
perturbations somewhat shift the local frame out of the Hubble flow
(in which $t$ would otherwise be the proper time
everywhere). Fortunately, there is a straightforward way to deal
with such `gauge' issues using the coordinate transformation 
\begin{equation}
{d\tau\over dt}=1-{\delta p\over \rho+p}\;.
\end{equation}

The energy conservation law reads
\begin{equation}
{d\rho\over d\tau}=-3H(\rho+p)\;,
\end{equation}
and it is straightforward to see from Equations~(5) and (6) that
(Padmanabhan 1993; Liddle and Lyth 2000)
\begin{equation}
{d(\delta\rho)\over dt}=-3H_0\,\delta\rho-3\,\delta H\,(\rho_0+p_0)\;.
\end{equation}
In addition,
\begin{equation}
{d\over dt}(\delta H)+2H_0\delta H+{4\pi G\over 3c^2}\rho_0\delta +
{v_s^2\over 3(1+w)}D^2\delta=0\;,
\end{equation}
and
\begin{equation}
\dot{H_0} = -{3\over 2}(1+w)H_0^2\;,
\end{equation}
where $H(t)=H_0(t)+\delta H(t)$, $H_0(t)$ is the smoothed Hubble constant
driven by the background fluid (i.e., $\rho_0$ and $p_0$) at time $t$,
$\delta H(t)$ describes scalar deviations from the smooth background
expansion rate represented by $H_0$, and $D^2$ is the 4-dimensional
Laplacian operator. ($H_0$ should not be confused with the Hubble constant
today, which is usually also denoted with subscript ``$0$".)

As we discuss below, $\rho_0$ in the $R_{\rm h}=ct$ universe appears
to be dominated by dark energy and (baryonic and dark) matter at low
redshifts, and dark energy and radiation in the early Universe. In this
paper, we focus on the more recent growth of perturbations, and we follow
the conventional approach of assuming that dark energy ($\rho_{\rm de}$)
is a smooth background, while the perturbations themselves are due solely
to fluctuations in the matter density $\rho_{\rm m}$. We further assume,
again conventionally, that once these matter perturbations start to
condense out of the smooth cosmic fluid, they decouple
from the dark energy, except through their gravitational interaction.

We shall see shortly that in the $R_{\rm h}=ct$ universe, where
$\rho+3p=0$, $\delta$ is constant to first order at low redshifts.
In solving Equations~(7)-(9), we will therefore also retain the leading
second-order terms to determine its time dependence, if any.
From Equation~(7), it is straightforward to show that
\begin{equation}
\dot{\delta}=-H_0\,\delta-2\,\delta H-3\,\delta H\,\delta\;,
\end{equation}
and therefore from Equations~(8)-(10), we find that
\begin{equation}
\ddot{\delta}+3H_0\,\dot{\delta}-{3\over 2}\left({\dot{\delta}}^2-
H_0\,\delta\,\dot{\delta}+H_0^2\,\delta^2\right)=v_s^2\,D^2\delta\;.
\end{equation}
To first-order, the gravitational (third) term on the left-hand side is
zero, a result of the zero-active mass condition $\rho+3p=0$, for
which the cosmic fluid experiences no net gravitational acceleration.
We therefore assume that $\delta$ is (at most) a very weak function
of $t$,
\begin{equation}
\delta(t)\sim t^\alpha\;,
\end{equation}
with $|\alpha|\ll 1$. In that case, 
\begin{equation}
{3\over 2}\left({\dot{\delta}}^2-H_0\,\delta\,\dot{\delta}+H_0^2\,\delta^2\right)=
{1\over t^2}\left(\alpha^2\delta^2-\alpha\delta^2+\delta^2\right)\;,
\end{equation}
which we approximate as
\begin{equation}
{3\over 2}\left({\dot{\delta}}^2-H_0\,\delta\,\dot{\delta}+H_0^2\,\delta^2\right)\approx
At^{-2}\,\delta\;,
\end{equation}
where $A$ is essentially constant with $|A|\ll 1$. And so Equation~(11) reduces to the form
\begin{equation}
\ddot{\delta}+3H_0\,\dot{\delta}-{A\over t^2}\delta=v_s^2\,D^2\delta\;.
\end{equation}

The derivative term in Equation~(15) requires some knowledge
concerning the spatial variation of the fluctuation. This is typically
handled via a wavenumber decomposition of $\delta\rho$, in which 
the density fluctuation is written as a Fourier series,
\begin{equation}
\delta(x^\alpha)={1\over (2\pi)^3}\int {\tilde{\delta}}_k(t)\,
e^{-i{\bf k}\cdot{\bf x}}\,d^3k\;,
\end{equation}
in terms of the comoving wavenumber $k$ and wavevector $k^\alpha\equiv
(\omega_k/c,{\bf k})$ where, as always, $k=|{\bf k}|$. The Fourier
components may be calculated through the expression
\begin{equation}
{\tilde{\delta}}_k(t)=\int\delta(x^\alpha)\,e^{i{\bf k}\cdot{\bf x}}\,d^3x\;.
\end{equation}
With this, and the fact that $H_0=1/t$,  it is straightforward to show from Equation~(15) 
that the local growth rate equation in the $R_{\rm h}=ct$ universe may be written as
\begin{equation}
{d^2{\tilde{\delta}}_k\over dt^2}+ {3\over t}{d{\tilde{\delta}}_k\over dt}-
{A\over t^2}{\tilde{\delta}}_k=
-{k^2\over a^2}v_s^2{\tilde{\delta}}_k\;,
\end{equation}
where $a(t)$ is the universal expansion factor. Note that for the
application we are considering in this paper, we have explicitly used
the constant value $w-1/3$, so terms proportional to $\dot{w}$ have
been omitted from these expressions (cf. Tsagas et al. 2008).

These equations specifically describe the evolution of scalar fluctuations.
It is well known that the inclusion of metric perturbations about the spatially
flat Friedmann-Robertson-Walker (FRW) background metric produces a combination
of modes that conveniently split into scalar, vector, and tensor components,
depending on how they transform on spatial hypersurfaces. But vector
perturbations have no lasting influence in an expanding universe. In
addition, scalar and tensor modes decouple to linear order, so gravity waves
do not provide any backreaction to the metric; they satisfy sourceless equations
when the energy-momentum tensor is diagonal, as is usually assumed for a perfect
fluid in cosmology. For these reasons, we focus exclusively on the
evolution of scalar perturbations in this paper.

\section{Perturbation Growth in the $R_{\rm h}=ct$ Universe}
\subsection{Background}
Let us first briefly discuss the motivation for considering this
model. We have been developing this Friedmann-Robertson-Walker cosmology 
for over ten years now (Melia 2007, 2016a, 2017; Melia \& Shevchuk 2012), 
driven largely by a series of observational tests that suggest 
its predicted expansion rate is a better fit to the data than that of the current
concordance model, $\Lambda$CDM. In this sub-section, we survey
some of these completed model comparisons, and explain why it is now
necessary to probe this cosmology more deeply, including its predicted
growth rate---the subject of this paper.

\begin{table}
  \caption{Model Comparisons between $R_{\rm h}=ct$ and $\Lambda$CDM}
  \centering
  \begin{tabular}{lll}
&& \\
    \hline
\hline
&& \\
Description & Outcome & Reference\\
&& \\
\hline
&& \\
Alcock-Paczy\'nski test with the BAO scale & $\Lambda$CDM is ruled out in comparison to $R_{\rm h}=ct$ at a $2.6\sigma$ c.l.&
Melia \& L\'opez-Corredoira (2016) \\
FSRQ $\gamma$-ray luminosity function& $R_{\rm h}=ct$ very strongly favoured over $\Lambda$CDM with $\Delta\gg10$ &Zeng et al. (2016) \\
QSO Hubble diagram $+$ Alcock-Paczy\'nski& $R_{\rm h}=ct$ $\sim 4$ times more likely than $\Lambda$CDM to be correct&
L\'opez-Corredoira et al. (2016) \\
Constancy of the cluster gas mass fraction & $R_{\rm h}=ct$ favoured over $\Lambda$CDM with BIC likelihood $95\%$ vs $5\%$&Melia (2016c) \\
Cosmic Chronometers&$R_{\rm h}=ct$ favoured over $\Lambda$CDM with BIC likelihood $95\%$ vs $5\%$&Melia \& Maier (2013); \\
&&Melia \& McClintock (2015a)\\
Cosmic age of old clusters&$\Lambda$CDM can't accommodate high-$z$ clusters, but $R_{\rm h}=ct$ can&Yu \& Wang (2014) \\
High-$z$ quasars&Evolution timeline fits within $R_{\rm h}=ct$, but not $\Lambda$CDM&Melia (2013); \\
&&Melia \& McClintock (2015b) \\
The AGN Hubble diagram& $R_{\rm h}=ct$ favoured over $\Lambda$CDM with BIC likelihood $96\%$ vs $4\%$&Melia (2015b) \\
Age vs. redshift of old passive galaxies& $R_{\rm h}=ct$ favoured over $\Lambda$CDM with BIC likelihood $80\%$ vs $20\%$&Wei et al. (2015a) \\
Type Ic superluminous supernovae & $R_{\rm h}=ct$ favoured over $\Lambda$CDM with BIC likelihood $80\%$ vs $20\%$&Wei et al. (2015b) \\
The SNLS Type Ia SNe&$R_{\rm h}=ct$ favoured over $\Lambda$CDM with BIC likelihood $90\%$ vs $10\%$&Wei et al. (2015c) \\
Angular size of galaxy clusters&$R_{\rm h}=ct$ favoured over $\Lambda$CDM with BIC likelihood $86\%$ vs $14\%$&Wei et al. (2015d) \\
Strong gravitational lensing galaxies&Both models fit the data very well due to the bulge-halo `conspiracy'& Melia et al. (2015c) \\
Time delay lenses&$R_{\rm h}=ct$ favoured over $\Lambda$CDM with BIC likelihood $80\%$ vs $20\%$&Wei et al. (2014a) \\
High-$z$ galaxies&Evolution timeline fits within $R_{\rm h}=ct$, but not $\Lambda$CDM&Melia (2014a) \\
GRBs $+$ star formation rate&$R_{\rm h}=ct$ favoured over $\Lambda$CDM with AIC likelihood $70\%$ vs $30\%$&Wei et al. (2014b) \\
High-$z$ quasar Hubble diagram&$R_{\rm h}=ct$ favoured over $\Lambda$CDM with BIC likelihood $85\%$ vs $15\%$&Melia (2014b) \\
GRB Hubble diagram&$R_{\rm h}=ct$ favoured over $\Lambda$CDM with BIC likelihood $96\%$ vs $4\%$&Wei et al. (2013) \\
&& \\
\hline\hline
  \end{tabular}
\end{table}

Since competing models tend to have different formulations, often with
unmatched parameters, one must use model selection tools to determine
which (if any) is preferred by the data. It is now common in cosmology to
use tools such as the Akaike Information Criterion (AIC; Liddle 2007), 
the Kullback Information Criterion (KIC; Cavanaugh 2004), and the
Bayes Information Criterion (BIC; Schwarz 1978) for this purpose. 
When using the AIC, with ${\rm AIC}_\alpha=-2\ln\mathcal{L}_\alpha+2n_\alpha$
characterizing model $\mathcal{M}_\alpha$, the difference $\Delta_{\rm AIC}=
{\rm AIC}_2-{\rm AIC}_1$ determines the extent to which model $\mathcal{M}_1$
is favoured over model $\mathcal{M}_2$. Here, $\mathcal{L}$ is the maximum
value of the likelihood function and $n$ is the number of free parameters 
(see Melia \& Maier 2013, and references cited therein, for more details).
For Kullback and Bayes, the likelihoods are defined analogously. In using
these model selection tools, the outcome $\Delta$ (for AIC, KIC, or BIC,
as the case may be) is judged to represent `positive' evidence that model
1 is preferred over model 2 if $\Delta>2$. If $2<\Delta<6$, the evidence
favouring model 1 is moderate, and it is very strong when $\Delta>10$.
Sometimes, the outcome $\Delta$ is used to estimate the relative probability
(or percentage likelihood) that $\mathcal{M}_1$ is statistically preferred
over $\mathcal{M}_2$, according to the prescription
\begin{equation}
P(\mathcal{M}_1) = {1\over 1+\exp{(-\Delta/2)}}\;,
\end{equation}
with $P(\mathcal{M}_2)=1-P(\mathcal{M}_1)$, when only two models
are being compared directly.

In Table 1, we quote the outcome $\Delta\equiv\Delta_{\Lambda{\rm CDM}}-
\Delta_{R_{\rm h}=ct}$, or the percentage likelihoods calculated from it,
for $R_{\rm h}=ct$ and $\Lambda$CDM, based on the 18 tests
published thus far. All of these outcomes have consistently
favoured $R_{\rm h}=ct$ over $\Lambda$CDM, sometimes moderately, 
often very strongly. Yet in spite of these consistently favourable
comparisons, some have been critical of the $R_{\rm h}=ct$ model. For
example, in contrast to the conclusions regarding the SNLS Type Ia 
SNe by Wei et al. (2015c), Shafer (2015) compared cosmological models
using both the Union2.1 (Suzuki et al. 2012) and JLA (Betoule et al.
2014) SN samples and argued that $\Lambda$CDM was strongly favoured
by these data. However, he appears to have incorrectly estimated
the intrinsic dispersion of each sub-sample, and additionally failed
to include them in his maximum likelihood estimation, which greatly
biased his analysis. Shafer (2015) also analyzed measurements of
$H(z)$ versus $z$, but here too he avoided using truly model-independent
cosmic chronometer measurements, opting instead to use heavily
biased BAO estimates. He appears to have been unaware of the 
significant limitations of all but the most recent 2 or 3 BAO 
scale determinations. A similar study, also based on model-dependent
estimates of $H(z)$ versus $z$, was carried out by Bilicki \& Seikel
(2012). In contrast, when one uses truly model-independent cosmic 
chronometer observations, the outcome strongly favours $R_{\rm h}=ct$ 
over $\Lambda$CDM (see, e.g., Melia \& Maier 2013; Melia \& 
McClintock 2015a).

Taking a different approach, van Oirschot et al. (2010)
and Lewis (2013) argued that the definition of the gravitational
horizon $R_{\rm h}$ in the $R_{\rm h}=ct$ universe is ill-defined
and that light rays emitted from beyond it are nonetheless still detectable.
But their analysis was flawed because it was based on an improper
use of null geodesics in the FRW metric. A full accounting of this, 
and a proof that no null geodesics reaching us today originated 
from beyond $R_{\rm h}$, appeared in Bikwa et al. (2012), and 
Melia (2012).

And in two of the more recent claims made against $R_{\rm h}=ct$,
Mitra (2014) has argued that this cosmology is static and merely 
represents another vacuum solution, while Lewis (2013) attempted
to show that the equation of state in $R_{\rm h}=ct$ is inconsistent
with $p=-\rho/3$, thereby ruining the elegant, high-quality fits
to the observations. As shown in Melia (2015c), however, these
criticisms are based on either incorrect assumptions or basic
theoretical errors. For example, Florides (1980) proved in his
landmark paper that there are six---and only six--- special cases
of the FRW metric for which one may transform the coordinates into
a frame where the metric coefficients are static. The $R_{\rm h}=ct$
model is not one of them. Mitra (2014) erroneously concluded that
this cosmology has a constant expansion rate because its density
is zero. In fact, the linear expansion occurs because $R_{\rm h}=ct$
has zero active mass, i.e., $\rho+3p=0$, not because $\rho=0$.
Lewis's (2013) analysis was even more superficial than this,
because he based his conclusion on the inexplicable assumption
that $\rho$ must have only a single component in the $R_{\rm h}=ct$
universe. Like $\Lambda$CDM, however, $\rho$ in this cosmology
has multiple components, including radiation, matter and the
poorly known dark energy. But these two models differ in the critical 
constraint that the total $p$ must always be equal to $-\rho/3$ in 
$R_{\rm h}=ct$, though not in $\Lambda$CDM. A more detailed explanation 
of how these components vary with redshift may be found in Melia \&
Fatuzzo (2016).

The growth rate described in this paper addresses 
one of the few remaining areas where an examination of how well these
two models account for the data has yet to be made. As we shall see,
our results indicate that the measured
growth rate favours $R_{\rm h}=ct$ over $\Lambda$CDM, in complete 
agreement with the outcomes presented in Table~1.
\vfill\newpage
\subsection{Growth Equation in the $R_{\rm h}=ct$ Universe}
Let us now apply the growth equation to the $R_{\rm h}=ct$ universe, in which
(like $\Lambda$CDM) the energy density contains at least five components:
(1) cold dark matter (cdm), (2) baryonic matter (b), (3) photons $(\gamma)$,
(4) neutrinos $(\nu)$, and (5) dark energy (de), so that generally
\begin{equation}
\rho=\rho_{\rm m}+\rho_{\rm r}+\rho_{\rm de}\;.
\end{equation}
The baryons and cold dark matter are often grouped together,
\begin{equation}
\rho_{\rm m}=\rho_{\rm b}+\rho_{\rm cdm}\;,
\end{equation}
as are the photons and neutrinos when the latter are still relativistic:
\begin{equation}
\rho_{\rm r}=\rho_\gamma+\rho_\nu\;.
\end{equation}
At least within the standard model ($\Lambda$CDM), baryons and photons
interact with each other up until the time of decoupling, $t_{\rm dec}$ so,
for $t<t_{\rm dec}$, they must be treated as a single component, and one
often writes
\begin{equation}
\rho_{{\rm b}\gamma}=\rho_{\rm b}+\rho_\gamma\;.
\end{equation}
Irrespective of which components may be dominant or interacting, however,
the perturbed Einstein Equations feature a single stress-energy tensor
representing the total energy density and pressure. In the standard model,
one must therefore make additional simplifying assumptions concerning some
of the components. One typically adopts the view that only one (or at most
two) components dominate the energy density and that the non-interacting
components affect the cosmic fluid only gravitationally. Of course,
a perfect fluid description can be applied to these components only
so long as their mean free paths are shorter than the scales of interest.
For example, after decoupling, photons stream freely and form a
homogeneous distribution that we can approximate as a `smooth'
background. The data also suggest that neutrino masses are probably
small enough to have only a minimal impact on structure formation, so
neutrinos too are usually approximated as a smooth radiation component.

These issues apply with equal validity to $R_{\rm h}=ct$, except
for one principal difference. In order for the cosmic fluid to maintain
a fixed equation-of-state $w\equiv p/\rho=-1/3$, dark energy cannot
be a cosmological constant (Melia 2015a; Melia and Fatuzzo 2016); it must
be dynamic, possibly with a particulate origin in physics beyond
the standard model. If we now follow a prescription similar to that
outlined in Equations~(20-23), then the simplest assumption we can
make (Melia and Fatuzzo 2016) is that dark energy and radiation (with a
contribution from both photons and neutrinos) dominated at early
times (i.e., $\rho\approx \rho_{\rm de}+\rho_{\rm r}$ for $z\gg 1$),
while dark energy and matter dominate our local universe (i.e.,
$\rho\approx \rho_{\rm de}+\rho_{\rm m}$ for small $z$). One can easily
show (Melia and Fatuzzo 2016) that for $z\gg 1$, this prescription would be
consistent with the partitioning
\begin{equation}
\rho_{\rm de}\approx {2\over 1-3w_{\rm de}}\rho_{\rm c}(1+z)^2\quad (z\gg 1)\;,
\end{equation}
and
\begin{equation}
\rho_{\rm r}\approx {3w_{\rm de}+1\over 3w_{\rm de}-1}\rho_{\rm c}(1+z)^2\quad (z\gg 1)\;,
\end{equation}
where $\rho_c \equiv {3 H_0(0)^2 / 8 \pi G}$ is the critical density.
Thus, if $w_{\rm de}$ were $-1/2$ towards higher redshifts, we would
have $\rho_{\rm de}\approx 0.8\rho$ and $\rho_{\rm r}\approx 0.2\rho$.
In this (perhaps over-simplified) scheme, one infers a gradual evolution
in the relative abundance of the various components, suggesting that they
remain coupled at all times. For $z\rightarrow 0$, radiation would no longer
have been dynamically important and $\rho$ would have been dominated by
$\rho_{\rm m}$ and $\rho_{\rm de}$. Again, it is straightforward to
show (Melia and Fatuzzo 2016) that, in order to maintain zero active mass
(i.e., $\rho+3p=0$),
\begin{equation}
{\rho_{\rm m}\over\rho}\approx 1+{1\over 3w_{\rm de}}\quad (z < 15)\;.
\end{equation}
For example, if $w_{\rm de}=-1/2$, this suggests that $\rho_{\rm m}/\rho\approx
1/3$ at low redshifts.

Quantum fluctuations emerging out of the Planck regime (Melia 2016b)
would have seeded perturbations in this background fluid.
A primordial scalar field $\phi$ with zero active mass, i.e., with an
equation of state $\rho_\phi+3p_\phi=0$, where $\rho_\phi$ and $p_\phi$
are its energy density and pressure, respectively, would have produced
an essentially scale-free fluctuation spectrum without inflation. This
mechanism is based on the Hollands-Wald concept of a minimum wavelength
for the emergence of quantum fluctuations into the semi-classical universe
(Hollands and Wald 2002). In this scenario, the $1^\circ-10^\circ$ fluctuations
in the cosmic microwave background (CMB) correspond almost exactly to the
Planck length at the time these modes were produced. In contrast to the
situation in $\Lambda$CDM, where the fluctuations transition back and forth
across the gravitational horizon, the fluctuations in $R_{\rm h}=ct$ have
a wavelength that grows in proportion to $R_{\rm h}$, and therefore the
perturbations in this cosmology grow in amplitude while their wavelength
remains a fixed fraction of the Hubble radius.

Current observations suggest that dark energy forms a smooth background
and does not condense. Thus, except in the late stages of clumping, when
matter would have collapsed completely to form stars and galaxies, most
of the perturbation growth would have been driven by quantum fluctuations
$\delta\rho=\delta\rho_\phi$ that transitioned first into $\delta\rho\sim
\delta\rho_{{\rm b}\gamma}$ (dominated by radiation, though with a
`contamination' of baryonic matter) in the early universe, followed by
another transition into $\delta\rho\sim\delta\rho_{\rm m}$ (dominated by
matter) at later times.

This is the framework we shall assume in using Equation~(18) to derive the
growth equation for matter perturbations when radiation is
relatively unimportant, i.e., when $\rho\approx \rho_{\rm m}+\rho_{\rm de}$.
Since in this case $v_s\approx 0$ for matter, we have
\begin{equation}
\ddot{\tilde{\delta}}_k+{3\over t}\,\dot{\tilde{\delta}}_k
-{A\over t^2}\tilde{\delta}_k=0\;,
\end{equation}
whose solution is the simple polynomial
\begin{equation}
\tilde{\delta}_k(t) = \left(C_1\, t^{-2} +C_2\, t^{A/2}\right)\;,
\end{equation}
where the coefficients $C_1$ and $C_2$ depend on initial conditions. Ignoring 
the inconsequential decaying mode, we therefore conclude that
\begin{equation}
{\tilde{\delta}}_k(t) \approx {\tilde{\delta}}_k(t_0)
\left({t\over t_0}\right)^{A/2}\;,
\end{equation}
where the quantity ${\tilde{\delta}}_k(t_0)$ is the $k$-mode amplitude
of the fluctuation today, i.e., at time $t_0$. And since $1+z=t_0/t$, we may also
write Equation~(29) as
\begin{equation}
{\tilde{\delta}}_k(z)\approx {\tilde{\delta}}_k(0)(1+z)^{-A/2}\;.
\end{equation}

\subsection{Variance of the Perturbations}
The perturbation ${\tilde{\delta}}_k(t)$ is usually assumed to be a Gaussian
random field, which means that the waves in the decomposition of Equation~(16)
have random phases. In this instance, the field may be specified entirely
by its power spectrum
\begin{eqnarray}
\qquad\qquad\qquad\qquad\qquad P({\bf k})&=&\langle{\tilde{\delta}}_k^*{\tilde{\delta}}_k\rangle\nonumber\\ 
&=&\langle|{\tilde{\delta}}_k|^2\rangle\;.
\end{eqnarray}
And for an isotropic distribution, the power spectrum, averaged over all
possible realizations, must be independent of direction:
\begin{equation}
P(k) = {1\over 4\pi}\oint \langle|{\tilde{\delta}}_k|^2\rangle\,d\Omega\;.
\end{equation}
The power spectrum may also be written as the Fourier transform of the
autocorrelation function using the Wiener-Khinchin theorem:
\begin{eqnarray}
\qquad\qquad\qquad\langle \delta^*({\bf x})\delta({\bf x}+{\bf y})\rangle&=&
\left\langle\int {d^3k^\prime\over (2\pi)^3}\int {d^3k\over (2\pi)^3}\,
{\tilde{\delta}}^*_{k^\prime}\,{\tilde{\delta}}_k\,\right.\times\nonumber\\
&\null&\left.\qquad\qquad
e^{i{\bf k}^\prime\cdot{\bf x}}\,e^{-i{\bf k}\cdot({\bf x}+{\bf y})}\right\rangle\nonumber\\
&=&\int {d^3k\over (2\pi)^3}\,P({\bf k})e^{-i{\bf k}\cdot{\bf y}}\;,
\end{eqnarray}
whose inversion gives
\begin{equation}
P({\bf k})=\int d{\bf y}\,\langle\delta^*({\bf x})\delta({\bf x}+{\bf y})\rangle
\,e^{i{\bf k}\cdot{\bf y}}\;.
\end{equation}
Using ${\bf y}$ as the axis about which the poloidal and azimuthal angles
are integrated to reduce the general expression to a single integral over $k$,
one may write Equation~(33) more conveniently as
\begin{equation}
\langle \delta^*({\bf x})\delta({\bf x}+{\bf y})\rangle=
4\pi\int{k^2\,dk\over (2\pi)^3}\,P(k){\sin ky\over ky}\;,
\end{equation}
where $y=|{\bf y}|$. The variance $\sigma$ of $\delta(x^\alpha)$ is given by
the autocorrelation function at ${\bf y}=0$:
\begin{equation}
\sigma^2=4\pi\int {k^2\,dk\over (2\pi)^3}\,P(k)\;.
\end{equation}

However, since the fluctuations $\delta({\bf x})$ exist on all spatial scales,
a more practical measure to use when comparing the power spectrum to the data
is the variance delimited within a specified volume. For this purpose, a window
function $W_R({\bf x})$ is introduced with a characteristic radius $R$, such
that $W_R$ is non-zero for $|{\bf x}| < R$ and decreases to zero for
$|{\bf x}|\gg R$. The perturbation is then replaced by the convolution integral
\begin{equation}
\delta_R(x^\alpha)\equiv\int \delta({\bf y},t)W_R(|{\bf x}-{\bf y}|)\,d^3y
\end{equation}
and, correspondingly, the power spectrum $P(k)$ must be replaced with
$P(k){\tilde{W}}_R^2(k)$, where ${\tilde{W}}_R$ is the Fourier transform
of $W_R({\bf x})$. The volume-delimited variance may thus be written
\begin{equation}
\sigma_R^2=4\pi\int {k^2\,dk\over (2\pi)^3}\,P(k)\,{\tilde{W}}_R^2(k)\;.
\end{equation}
For a conventional Gaussian window,
\begin{equation}
W_R(y)={1\over (2\pi)^{3/2}R^3}\,e^{-y^2/2R^2}\;,
\end{equation}
the Fourier transform is
\begin{equation}
{\tilde{W}}_R(k) = e^{-(kR)^2/2}\;,
\end{equation}
so putting together Equations~(31) and (38-40), we arrive at the final expression for $\sigma_R$
in the $R_{\rm h}=ct$ universe:
\begin{equation}
\sigma_R^2(z)=4\pi\int {k^2\,dk\over (2\pi)^3}\,\langle|\tilde{\delta}_k(z)|^2\rangle \,e^{-(kR)^2}\;.
\end{equation}
By convention, this variance is usually calculated in spherical volumes with
a radius of 8 $h^{-1}$ Mpc. Thus, defining
\begin{equation}
\sigma_R^2(0)\equiv 4\pi\int {k^2\,dk\over (2\pi)^3}\,\langle|\tilde{\delta}_k(0)|^2\rangle \,e^{-(kR)^2}\;,
\end{equation}
the volume-delimited variance of
$\delta(x^\alpha)$ in the $R_{\rm h}=ct$ universe is given by the simple expression:
\begin{equation}
\sigma_8^{R_{\rm h}=ct}(z) \approx \sigma_8^{R_{\rm h}=ct}(0)(1+z)^{-A/2}\;,
\end{equation}
and since $|A|\ll 1$, we have for the low-redshift limit
\begin{equation}
\sigma_8^{R_{\rm h}=ct}(z) \approx \sigma_8^{R_{\rm h}=ct}(0)\;.
\end{equation}
This expression is valid as long as $v_s\approx 0$, i.e., as long as $\delta$ is
primarily a fluctuation of matter decoupled (other than through gravity) from 
the smooth dark energy background. It breaks down when $\rho$ includes 
a non negligible contribution from radiation.

\subsection{Observables in the $R_{\rm h}=ct$ Universe}
Measuring the growth rate of cosmological density perturbations is 
a promising method of testing cosmological models, given that
$\tilde{\delta}_k(z)$ may in some cases depend sensitively on the underlying
expansion rate $H(z)$  and, therefore, on
the equation of state $p=w\rho$. The evolution in the variance
$\sigma_8^{R_{\rm h}=ct}(z)$ is a manifestation of this growth rate, but galaxies form
only in the densest regions of the Universe, so their observed distribution
$\delta_{\rm g}$ is related to the matter density perturbations
$\delta$ via a non-trivial bias factor $b$: $\delta_{\rm g}=
b\,\delta$. Unfortunately, this bias varies between
different populations of galaxies, so measurements of $\sigma_8$ from
different surveys are difficult to combine and compare with theoretical
predictions.

An alternative approach is based on the measurement of peculiar
velocities from redshift space distortions in a galaxy redshift
survey, as first proposed by Kaiser (1987). These peculiar
velocities represent (small) deviations from a pure Hubble flow,
and are proportional to the so-called cosmological growth factor
\begin{equation}
f(a)\equiv {d\,\ln D(a)\over d\,\ln a}\;,
\end{equation}
where $D(a)$ represents the growth of matter fluctuations, defined
by the expression
\begin{equation}
\tilde{\delta}_k(a)=\tilde{\delta}_k(1)\,D(a)\;.
\end{equation}
In the $R_{\rm h}=ct$ universe, $a(t)=t/t_0$, so
\begin{equation}
D^{R_{\rm h}=ct}(a)= a^{A/2}\;,
\end{equation}
which means that, in this cosmology, $f^{R_{\rm h}=ct}(a)$ or, 
equivalently, $f^{R_{\rm h}=ct}(z)$, is constant in the same
redshift range where $\sigma_8$ is a very weak function of $z$.

\vskip 0.1in
\begin{table}
  \caption{Measurement of $f\sigma_8(z)$ from various redshift surveys}
  \centering
  \begin{tabular}{lllr}
&& \\
    \hline
\hline
&& \\
$\,\;\;z$ & $\;f(z)\sigma_8(z)$ & \quad\;\; Survey & Reference\\
&& \\
\hline
&& \\
$0.067$ & $0.42\pm0.05$ & 6dFGRS(2012) & (Jones et al. 2009; Beutler et al. 2012) \\
$0.22$  & $0.42\pm0.07$ & WiggleZ(2011) & (Blake et al. 2011) \\
$0.25$  & $0.35\pm0.06$ & SDSS LRG(2011) & (Eisenstein et al. 2011) \\
$0.37$  & $0.46\pm0.04$ & SDSS LRG(2011) & (Eisenstein et al. 2011) \\
$0.41$  & $0.45\pm0.04$ & WiggleZ(2011) & (Blake et al. 2011) \\
$0.57$  & $0.462\pm0.041$ & BOSS CMASS & (Dawson et al. 2013 Alam et al. 2015) \\
$0.60$  & $0.43\pm0.04$ & WiggleZ(2011) & (Blake et al. 2011) \\
$0.78$  & $0.38\pm0.04$ & WiggleZ(2011) & (Blake et a. 2011) \\
$0.80$  & $0.47\pm0.08$ & Vipers(2013) & (de la Torre et al. 2013) \\
&& \\
\hline\hline
  \end{tabular}
\end{table}

Over the past several decades, $f(z)$ has been measured using a range of techniques
and surveys, including 2dFGRS (Peacock et al. 2001), VVDS (Guzzo et al. 2008), quasar
clustering and Ly$_\alpha$ clustering (Ross et al. 2007; da Angela et al. 2008; Viel et al. 2004),
and in peculiar velocity surveys at $z\sim 0$ (Davis et al. 2011; Hudson et al. 2012).
For example, growth rate measurements may be made using the galaxy two-point
correlation function, which yields a value of the parameter $\beta\equiv f/b$.
Thus, here too the measured value of $f(z)$ is known from the galaxy distribution
only to within a bias factor $b$. Clearly, the probative power of both $\sigma_8(z)$
and $f(z)$ is mitigated by the uncertainty in this inferred bias. But each has a
dependence on $b$ that is the inverse of the other, so the product $f(z)\sigma_8(z)$
may be a much more suitable measure of the structural evolution
(Percival et al. 2009; Macaulay et al. 2013; Pavlov et al. 2014; Alam et al. 2016). 
Measurements of the growth
rate are now commonly reported using the quantity $f(z)\sigma_8(z)$, and these
are the data we will employ for the analysis in this paper (see Table 2).

Care must be taken with the use of these data, however, because the 
measurements are not all completely independent (Alam et al. 2016). Though
based on six different (mostly independent) surveys, in some cases probing 
different biased tracers, they are nonetheless essentially diagnosing the same 
matter density field, so some of the sampled volumes overlap. Alam et al. 
(2016) have calculated the fractional overlap volume between each pair of 
samples, from which they then estimated the correlation between the 
corresponding measurements (see their figure~2). 

For our model comparisons, which are based exclusively on the data listed
in Table 2 (and, correspondingly, the data in Table 3 recalibrated for 
$R_{\rm h}=ct$), we use maximum likelihood estimation with a likelihood function
\begin{equation}
\mathcal{L} \propto e^{-\chi^2/2}\;,
\end{equation}
where 
\begin{equation}
\chi^2=\Delta^T\,C^{-1}\,\Delta\;.
\end{equation}
Here, $C^{-1}$ is the inverse of the covariance matrix calculated from the 
measurement errors quoted in Table 2 and the correlation matrix, and $\Delta$ 
is the column vector expressing the differences between the measured and
predicted values of $f\sigma_8$. Its individual components are
\begin{equation}
\Delta_i\equiv f\sigma_8(z_i)|_{\rm theory}-f\sigma_8(z_i)|_{\rm obs}\;.
\end{equation}

But before we can test the model, we must address an additional complication.
The galaxy surveys do not measure distances directly. To convert from a measured
redshift to a physical distance, one must pre-assume a cosmological model. The
problem, of course, is that additional (artificial) anisotropies due to the
Alcock-Paczy\'nski effect (Alcock and Paczy\'nski 1979; Melia and L\'opez-Corredoira 
2016) may be imposed on
the redshift space distortions if the chosen model is incorrect. One must
therefore recalibrate the data for each model being tested. All of the entries
in Table~2, with the exception of the BOSS datum, were obtained assuming a
fiducial $\Lambda$CDM cosmology with WMAP optimized parameter values (Hinshaw et al. 2013).
The BOSS measurement was made with the Planck concordance model as background
(Planck Collaboration 2014b).

Fortunately, there is a way of transforming the 3-dimensional two-point
correlation function from one model to another using the Alcock-Paczy\'nski
effect (Macaulay et al. 2013). Using ``fid" to designate parameter values
optimized with the pre-assumed fiducial model, one may obtain the corresponding
quantities in the model being tested using the approximate formula
\begin{equation}
\left[f\sigma_8(z)\right]^{R_{\rm h}=ct}=\left[f\sigma_8(z)\right]^{\rm fid}\times B\;,
\end{equation}
where
\begin{equation}
B\equiv {H^{\rm fid}(z)\over H^{R_{\rm h}=ct}(z)}{d_A^{\rm fid}(z)\over d_A^{R_{\rm h}=ct}(z)}\;.
\end{equation}
In this expression, $H(z)$ is the redshift-dependent Hubble constant,
and $d_A$ is the angular-diameter distance. For the application we are
considering in this paper, we have
\begin{equation}
H^{R_{\rm h}=ct}(z) = H(0)(1+z)\;,
\end{equation}
\begin{equation}
H^{\rm fid}(z) = H(0)\left[\Omega_{\rm m}(1+z)^3+(1-\Omega_{\rm m}-\Omega_\Lambda)(1+z)^2
+\Omega_\Lambda\right]^{1/2}\;,
\end{equation}
and
\begin{equation}
d_A^{R_{\rm h}=ct}(z)={c\over H(0)}{1\over (1+z)}\ln(1+z)\;,
\end{equation}
\begin{equation}
d_A^{\rm fid}(z)={c\over H(0)}{1\over (1+z)}\int_0^zdu{H(0)\over H(u)}\;.
\end{equation}
Note that we are here using the symbol $H(0)$ to represent the Hubble constant
today in order to avoid confusion with our previously defined $H_0$, while $\Omega_i\equiv
\rho_i/\rho_c$ is the current fractional energy density of species ``$i$" in
terms of the critical density $\rho_c\equiv 3c^2H_0^2/8\pi G$.
The measured values of $\left[f\sigma_8(z)\right]^{R_{\rm h}=ct}$ for the 
$R_{\rm h}=ct$ cosmology, recalibrated using this procedure, are listed in Table 3.

\vskip 0.1in
\begin{table}
  \caption{Observed values of $f\sigma_8(z)$ recalibrated using Equation~(51) for the $R_{\rm h}=ct$ cosmology}
  \centering
  \begin{tabular}{lllr}
&& \\
    \hline
\hline
&& \\
$\,\;\;z$ & $\;f(z)\sigma_8(z)$ & \quad\;\; Survey & Reference\\
&& \\
\hline
&& \\
$0.067$ & $0.41\pm0.05$ & 6dFGRS(2012) & (Jones et al. 2009; Beutler et al. 2012) \\
$0.22$  & $0.40\pm0.07$ & WiggleZ(2011) & (Blake et al. 2011) \\
$0.25$  & $0.33\pm0.06$ & SDSS LRG(2011) & (Eisenstein et al. 2011) \\
$0.37$  & $0.43\pm0.04$ & SDSS LRG(2011) & (Eisenstein et al. 2011) \\
$0.41$  & $0.42\pm0.04$ & WiggleZ(2011) & (Blake et al. 2011) \\
$0.57$  & $0.44\pm0.04$ & BOSS CMASS & (Dawson et al. 2013; Alam et al. 2015) \\
$0.60$  & $0.40\pm0.04$ & WiggleZ(2011) & (Blake et al. 2011) \\
$0.78$  & $0.36\pm0.04$ & WiggleZ(2011) & (Blake et al. 2011) \\
$0.80$  & $0.44\pm0.08$ & Vipers(2013) & (de la Torre et al. 2013) \\
&& \\
\hline\hline
  \end{tabular}
\end{table}

\section{Perturbation Growth in $\Lambda$CDM}
The linear growth equation in $\Lambda$CDM, which may also be derived from
the formalism in \S~2, is well known and we simply adopt the key results from previous
work (e.g., Linder 2005). It is often written in the form
\begin{equation}
{\ddot{\delta}}_k+2H{\dot{\delta}}_k={4\pi G\over c^2}\rho\,\delta_k
\end{equation}
where, as usual, dot indicates a derivative with respect to cosmic time. 
The Hubble constant $H(z)$ is defined in Equation~(54)
and, in the case of $\Lambda$CDM, $\Omega_\Lambda$ is assumed to be a cosmological constant 
(for which the dark-energy pressure is $p_{\rm de}=-\rho_\Lambda$). In addition, for flat $\Lambda$CDM
(the fiducial model used here), $\Omega_k=0$. We have adopted the Planck optimized values (Planck
Collaboration 2014a), for which $H(0)=67.4\pm1.4$ km s$^{-1}$ Mpc$^{-1}$, $\Omega_m=0.314\pm0.020$,
and $\Omega_\Lambda=0.686\pm0.020$. 

The solution to Equation~(57) may be written
\begin{equation}
\delta_k\propto D^{\Lambda{\rm CDM}}(z) = D_0 H^{\rm fid}(z)\int_z^\infty
{1+z^\prime\over H^{\rm fid}(z^\prime)^3}\,dz^\prime\;,
\end{equation}
where $D_0$ is a normalization constant. Therefore, 
\begin{equation}
\sigma_8^{\Lambda{\rm CDM}}(z)=\sigma_8^{\Lambda{\rm CDM}}(0)D^{\Lambda{\rm CDM}}(z)\;,
\end{equation}
normalized such that $D^{\Lambda{\rm CDM}}(0)=1$. In addition, we may calculate
the growth factor in this model using
\begin{equation}
f^{\Lambda{\rm CDM}}(z) = {\ddot{a}a\over{\dot{a}}^2}-1+{5\Omega_{\rm m}\over 2}
{(1+z)^2H_0^2\over H^{\rm fid}(z)^2D^{\Lambda{\rm CDM}}(z)}\;.
\end{equation}

The clumping of matter in $\Lambda$CDM is being studied at both high redshifts,
primarily with the analysis of anisotropies in the cosmic microwave background (CMB;
see, e.g., Planck Collaboration 2014a, 2014b), and at low redshifts, with weak
lensing, galaxy clustering, and the abundance of galaxy clusters (see, e.g., MacCrann
et al. 2015). The current status of this work points to some tension between the
predictions of $\Lambda$CDM and the measured linear growth rate at low and
high redshifts. The primary CMB anisotropies place limits on the matter
fluctuation amplitude at the time of recombination that may be extrapolated to
the nearby universe. But the low-redshift measurements seem to be finding
a lower value for this fluctuation amplitude than is expected in $\Lambda$CDM
(Vikhlinin et al. 2009; Planck Collaboration 2014a, 2014b; Beutler et al. 2014;
MacCrann et al. 2015). 

\begin{figure}
\centerline{\includegraphics[angle=0,scale=0.8]{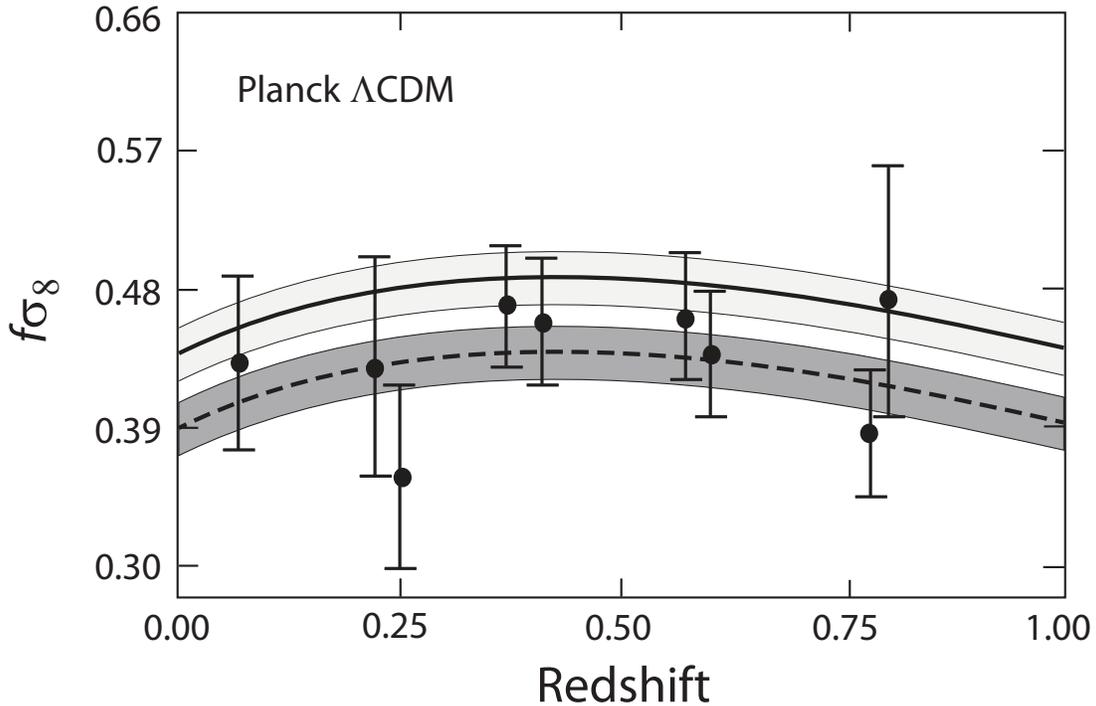}}
\caption{Measured values of $f\sigma_8(z)$ versus redshift from various surveys,
as indicated in Table 2. The data plotted here have all been recalibrated for
the Planck cosmology (Planck Collaboration 2014a). The solid curve shows the Planck 
$\Lambda$CDM prediction, based on the joint analysis of the {\it Planck} measurements
and the growth rate measurements listed in Table~2, with the shaded (light-gray) band
giving the 1 $\sigma$ confidence region (adapted from Alam et al. 2016). The
$\chi^2$ for this fit is 8.4 with $9-3=6$ degrees of freedom. The dashed curve, and 
its corresponding 1 $\sigma$ (dark-gray) confidence region, shows the $\Lambda$CDM 
fit based solely on the linear growth data in Table 2. The $\chi^2$ for this fit 
is 4.03 with $9-1=8$ degrees of freedom (though with the adoption of two
priors).}\label{LCDM}
\end{figure}

We can see this effect directly by comparing the two theoretical curves
superimposed on the fiducial data in figure~1. The solid curve, with its
associated (light-gray) 1 $\sigma$ confidence region, shows the optimized
$\Lambda$CDM fit based on the joint analysis of the {\it Planck} and linear
growth rate data (adapted from Alam et al. 2016). The $\chi^2$ for this
fit is 8.4, with $9-3=6$ degrees of freedom. By comparison, the dashed
curve (and associated dark-gray 1 $\sigma$ confidence region) shows
$\Lambda$CDM's best fit based solely on the growth rate
data in Table~2. The $\chi^2$ for this fit is $4.03$ with $9-1=8$
degrees of freedom. The principal difference between these two
curves is the value of $\sigma_8(0)$. The latter optimization has
a local fluctuation amplitude approximately $10\%$ smaller than
that of the former, which is dominated by the value of $\sigma_8(z)$
at recombination. We will rejoin this discussion shortly, following 
our analysis of $f\sigma_8(z)$ for the $R_{\rm h}=ct$ cosmology.

\section{Discussion}
The data recalibrated using Equation~(51) for the $R_{\rm  h}=ct$ cosmology
are shown in figure~2, together with the optimized theoretical fit in this
model. For the purpose of this analysis, $\sigma_8(0)$ is the sole free
parameter that may be adjusted in the fitting procedure. Also shown in
figure~2 is the 1 $\sigma$ confidence region, corresponding to the
best-fit value $f\sigma_8(0)=0.40\pm0.03$, with a total $\chi^2$ of $4.8$
and $9-1=8$ degrees of freedom. Clearly, the $R_{\rm h}=ct$ cosmology
fits the linear growth-rate data very well, arguably even better than the
version of $\Lambda$CDM optimized to fit both the {\it Planck} and growth rate
data (fig.~1, solid curve; adapted from Alam et al. 2016). We stress, however, 
that when $\Lambda$CDM is optimized to fit solely the linear growth rate data 
(Table~2; fig.~1, dashed curve), the quality of the $R_{\rm h}=ct$ fit is statistically
indistinguishable from that of $\Lambda$CDM when two of its free parameters
are assumed to have prior values.

In future work, it will be essential to examine how the growth
rate in $R_{\rm h}=ct$ impacts our interpretation of the CMB anisotropies,
particularly with regard to the implied value of $\sigma_8^{R_{\rm h}=ct}(z)$
at recombination, and a comparison of its extrapolated value with 
$\sigma_8^{R_{\rm h}=ct}(0)$ measured locally. Unfortunately, the 
value of $\sigma_8^{\Lambda{\rm CDM}}(0)$ optimized for $\Lambda$CDM 
does not apply to $R_{\rm h}=ct$, whose angular diameter distance
and the ratio $\rho_{\rm m}/\rho$ at high redshift are quite different
from their counterparts in $\Lambda$CDM (see, e.g., Melia \& Shevchuk
2012; Melia \& Fatuzzo 2016). The value of $\sigma_8^{R_{\rm h}=ct}(0)$
presented in this paper is in fact the first (and, so far, only) evaluation of 
this fluctuation amplitude in the context of $R_{\rm h}=ct$.

\begin{figure}
\centerline{\includegraphics[angle=0,scale=0.9]{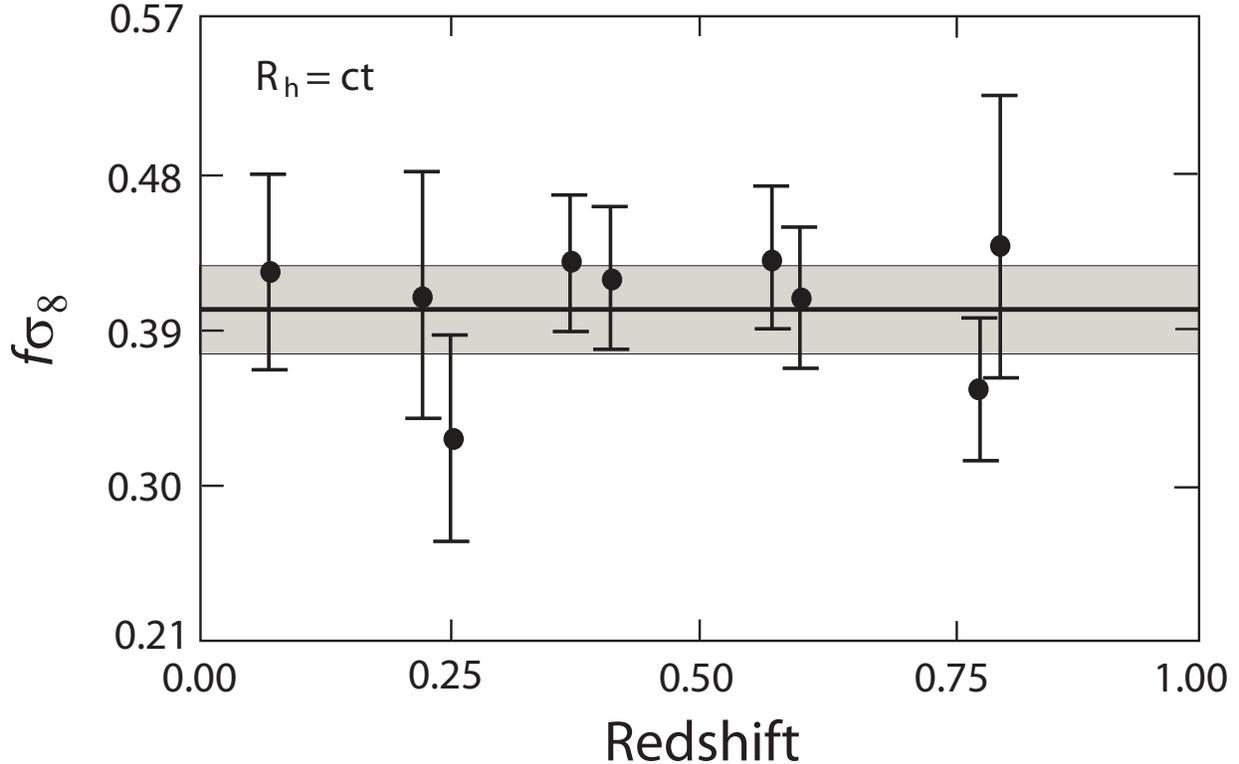}}
\caption{Same as fig.~1, except the data here have been recalibrated
using Equation~(51) for the $R_{\rm h}=ct$ universe, as indicated in
Table 3. The solid curve shows the best-fit in this cosmology, and
the shaded region is the 1 $\sigma$ confidence region, corresponding
to the optimized value $f\sigma_8(0)=0.40\pm0.03$. The $\chi^2$
for this fit is 4.3, with $9-1=8$ degrees of freedom.}\label{Rhct}
\end{figure}

Given these limitations, and fully acknowledging this important caveat,
we will for now directly compare the growth rate in $R_{\rm h}=ct$
measured locally with that implied by the version of $\Lambda$CDM 
optimized to fit both the low- and high-redshift data. It is quite evident
that the quality of the fit is not the only indication that $R_{\rm h}=ct$
may be preferred by these data. Even a simple inspection by eye would
suggest that these measurements, at least as shown in figure~2, point to
an absence of significant curvature in the measured functional dependence of
$f\sigma_8(z)$ on redshift. This is borne out by figure~3, which
compares the residuals in $\Lambda$CDM with those in $R_{\rm h}=ct$.
Notice, in particular that, whereas 6 out of the 9 measurements in
$R_{\rm h}=ct$ lie within 1 $\sigma$ of the best-fit curve, only 4
do so in $\Lambda$CDM. Worse, all 5 of the remaining points in the
standard model lie below the best-fit curve, while a purely randomized
distribution should have been evenly dispersed above and below it.
At the moment, this asymmetry in the $\Lambda$CDM residuals is an even
more compelling argument in favour of $R_{\rm  h}=ct$ than a simple
comparison of their $\chi^2$ values.

Unfortunately, one cannot be more definitive than this because,
in spite of the evident superiority of $R_{\rm h}=ct$ over
$\Lambda$CDM based on their residuals (and, to some degree, on
the quality of the fits), the growth rate measurements are not
yet accurate enough for us to clearly distinguish a model whose
best-fit curve has significant curvature (fig.~1) from one that
does not (fig.~2).

\section{Conclusion}
In recent years, perturbation theory has matured to the point
where the predictions of $\Lambda$CDM have been compared extensively
to measurements of the growth rate $f\sigma_8(z)$, and to other models.
The current consensus is that Planck $\Lambda$CDM (Planck Collaboration 2014a)
is generally well matched to the data, and that no firm evidence exists for extensions
to general relativity (Alam et al. 2016). Several recent analyses, however,
have yielded some tension between the value of $\sigma_8(0)$ measured
using redshift space distortions in galaxy surveys and that inferred
by fitting anisotropies in the cosmic microwave radiation
(Guzzo et al. 2008; Macaulay et al. 2013).

\begin{figure}
\centerline{\includegraphics[angle=0,scale=0.7]{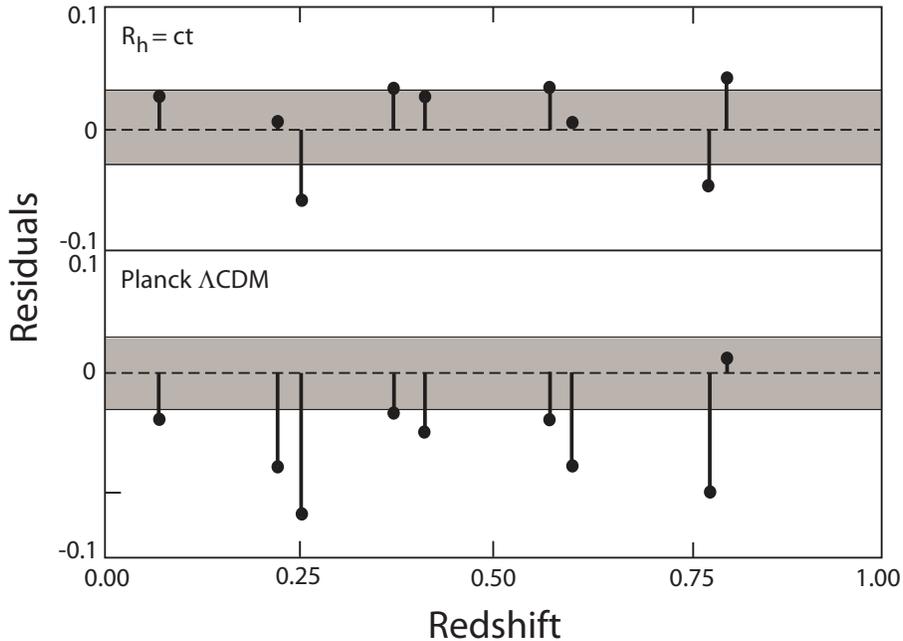}}
\caption{Residuals relative to the best-fit (solid) curves in figures~1 and 2.
Shaded sections correspond to 1 $\sigma$ confidence regions. For $\Lambda$CDM,
only 4 of the 9 points lie within 1 $\sigma$ of the optimized model. More
critically, all 5 of the remaining points lie below it. Together with the
indication given by the $\chi^2$ values, these results suggest that the linear
growth-rate measurements favour $R_{\rm h}=ct$ over Planck $\Lambda$CDM
(Planck Collaboration 2014a).}\label{Residuals}
\end{figure}

The principal goal of this paper has been to ascertain whether or not
the predictions of the $R_{\rm h}=ct$ universe are also consistent with
the measured growth rate at redshifts $z < 2-3$. We have found
that the current measurements, though still not sufficiently precise
to clearly distinguish between $R_{\rm h}=ct$ and $\Lambda$CDM,
nonetheless favour the former over the version of the latter
optimized by the joint analysis of {\it Planck} and linear growth
rate data. The two models are statistically indistinguishable when
the optimization of the $\Lambda$CDM fit is based solely on the growth
rate data in Table 2. Our results also suggest
that the present consistency of the standard model with the growth-rate
data may be an artifact of the relatively large errors associated
with these measurements, which cannot yet clearly distinguish between
functional forms of $f\sigma_8(z)$ with and without significant
curvature. This work strongly affirms the need for more precise measurements
of the growth rate in that critical redshift range ($0 < z < 1$)
where differences in the growth function $f\sigma_8(z)$ between
$\Lambda$CDM and $R_{\rm h}=ct$ are most pronounced.

\section*{Acknowledgments}
It is a pleasure to acknowledge helpful discussions with Christos Tsagas.
I am also very grateful to the anonymous referee for his thoughtful and insightful 
comments. These have led to a significant improvement in the presentation of
the paper. Some of this work was carried out at Purple Mountain Observatory in Nanjing,
China, and  was partially supported by grant 2012T1J0011 from The Chinese
Academy of Sciences Visiting Professorships for Senior International Scientists.

\end{document}